\documentclass[10pt,a4paper]{article}
\usepackage{amsmath,amssymb,graphicx,caption,amsthm}
\newtheorem* {MainThm}{Main Theorem}

\newtheorem{Lem}{Lemma}
\newtheorem{Rem}{Remark}

\begin{document}
\title{Convergent discrete Laplace-Beltrami operators over surfaces}
\author{Jyh-Yang Wu \thanks{Department of Mathematics, National Chung Cheng University, Chia-Yi 621,
Taiwan. Email:jywu@math.ccu.edu.tw}, Mei-Hsiu Chi\thanks{Department of Mathematics, National Chung Cheng University, Chia-Yi 621,
Taiwan. Email:mhchi@math.ccu.edu.tw} and Sheng-Gwo Chen \thanks{Department of Applied Mathematics, National Chiayi University,  Chia-Yi 600,
Taiwan. Email:csg@mail.ncyu.edu.tw (corresponding author)} }

\date{}
\maketitle
\begin{abstract}
The convergence problem of the Laplace-Beltrami operators plays an
essential role in the convergence analysis of the numerical
simulations of some important geometric partial differential
equations which involve the operator. In this note we present
a new effective and convergent algorithm to compute discrete
Laplace-Beltrami operators acting on functions over surfaces. We
prove a convergence theorem for our discretization. To our
knowledge, this is the first convergent algorithm of discrete
Laplace-Beltrami operators over surfaces for functions on general
surfaces. Our algorithm is conceptually simple and easy to compute.
Indeed, the convergence rate of our new algorithm of discrete
Laplace-Beltrami operators over surfaces is $O(r)$ where r
represents the size of the mesh of discretization of the surface.

\end{abstract}


\section{Introduction}
Let $\Sigma$ be a smooth surface in the 3D space. The
Laplace-Betrami (LB) operator is a natural generalization of the
classical Laplacian  $\Delta$ from the Euclidean space to $\Sigma$.
It is well-known that the LB operator is closed related to the mean
curvature normal by the relation $\Delta_{\Sigma}(p) =2H(p)$. The LB operator
plays important role not just in the study of geometric properties
of $\Sigma$, but also in the investigation of physical problems,
like heat flow and wave equations, on $\Sigma$. Moreover, the LB
operator has recently many applications in a variety of different
areas, such as surface processing \cite{Clarenz,Rusinkiewicz},
signal processing \cite{Sapiro, Schneider, Taubin, Taubin3} and geometric partial
differential equations \cite{Bertalmio,Theisel,Romeny}. Since the
objective underlying surfaces to be considered are usually
represented as discrete meshes in these applications, there are
tremendous needs in practice to discretize the LB operators.

Even though the computation of the LB operators is important for
many applications, there does not exist a simple "convergent" discrete
approximation of the LB operators for general surfaces. In this
paper we shall present a new effective and convergent algorithm to
compute discrete Laplace-Beltrami operators acting on functions over
surfaces. In fact, we shall prove the following convergence theorem.

\begin{MainThm}
Given a smooth function $h$ on a regular surface $\Sigma$  and a triangular surface mesh $S=(V,F)$  with mesh size $r$, one has
\begin{equation}
\Delta_{\Sigma} h(v) = \Delta_A h(v) +O(r)
\end{equation}
where the discrete LB operator $\Delta_A h(v)$ is given in Equation (\ref{laplace_h}).
\end{MainThm}

We shall give a mathematical proof of this convergence result. To our knowledge, this is the first convergent
algorithm of discrete Laplace-Beltrami operators over surfaces for
functions on general surfaces. The idea of our algorithm can be divided into two parts: First, we shall introduce a notation of the
local tangential polygon and lift functions and vectors on a
triangular mesh, obtained from the discretization of the surface
under consideration, to the local tangential polygon, and second, we shall
give a new method to define the discrete Laplace-Beltrami(LB)
operator acting on functions on a 2D polygon. Our algorithm is
conceptually simple and easy to compute. The convergence
rate of our new algorithm of discrete Laplace-Beltrami operators
over surfaces is $O(r)$ where r represents the size of the mesh of
discretization of the surface. We also present our numerical results
to support this in section 4.

\section{ Laplace-Beltrami operator and its discretizations}

Let $\Sigma$ be a regular surface in the 3D Euclidean space
$\mathbb{R}^3$. Consider a local parameterization $h : U \rightarrow
\Sigma$ with  $h(u_1, u_2) = ( x(u_1,u_2),y(u_1,u_2),z(u_1,u_2)) \in
\Sigma$, $(u_1, u_2) \in U \subset \mathbb{R}^2$. For the details,
we refer to do Carmo \cite{Docarmo, Docarmo2}. Then the Laplace-Beltrami
operator $\Delta_{\Sigma}$ applying to a $C^2$  function $f$  on $\Sigma$ is
given by
\begin{equation}
\Delta_{\Sigma} f = \frac{1}{\sqrt{g}}\sum_{ij}\frac{\partial}{\partial u_i}
\left [ g^{ij} \sqrt{g} \frac{\partial f}{\partial j} \right ]
\end{equation}
where $\begin{pmatrix} g^{ij} \end{pmatrix}$  is the inverse of the matrix
$\begin{pmatrix} g_{ij} \end{pmatrix}$  with $g = \det
\begin{pmatrix} g_{ij}\end{pmatrix}$ and
\begin{equation}
g_{ij} = < \frac{\partial h}{\partial u_i}, \frac{\partial
h}{\partial u_j}>.
\end{equation}

Consider a triangular discretization  $S = (V,F)$ of the surface
$\Sigma$, where $V = \{ v_i | 1 \leq i \leq n_V \}$ is the list of
vertices and $F = \{ T_k | 1 \leq k \leq n_F\}$ is the list of
triangles. Let $v$ be a vertex in $V$ and $N(v)$ the index of
one-ring neighbors of the vertex $v$ . Next we recall several
discretizations of $\Delta f$ for a $C^2$  function $f$ on as
follows. For more discussions, see also Xu\cite{Xu, Xu2}.

\subsection{Taubin's et al. Discretization}
\begin{figure}[htb]
{\center
\includegraphics[width=6cm]{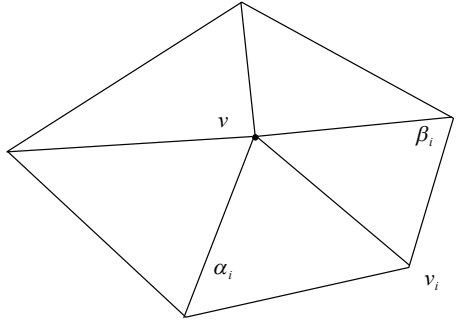}
\caption{The angles $\alpha _i$ and $\beta _i$.}\label{figure_angle}
}
\end{figure}

Taubin considered in \cite{Taubin} the following form of
discretization of $\Delta f$:
\begin{equation}\label{weight_method}
\Delta f(v) = \sum_{ i \in N(v)} \omega_{i} (f(v_i) - f(v))
\end{equation}
where the weights $\omega_i$ are nonnegative numbers with
$\sum_{i\in N(v)} \omega_i = 1$. There are several choices for
the weights $\omega_i$. An obvious choice is the uniform weights
$\omega_i = \frac{1}{|N(v)|}$ where $|N(v)|$ is the cardinality
of the set $N(v)$. A general way to determine the weights
$\omega_i$ is to use the following formulation:
\begin{equation}
\omega_i = \frac{\phi(v,v_i)}{\sum_{k \in N(v)} \phi(v,v_k)}
\end{equation}
with a nonnegative function $\phi(v,v_i)$. Fujiwara takes
$\phi(v,v_i) = \frac{1}{\| v_i-v\|}$. Desbrun's et al.
\cite{Desbrun} defines the weights $\omega_i$ as

\begin{equation}\label{Desbrun_weight}
\omega_i = \frac{\cot \alpha_i + \cot \beta_i}{\sum_{k \in
N(v)} \cot \alpha_k + \cot \beta_k}
\end{equation}
where $\alpha_{i}$ and $\beta_{i}$  are the triangles as shown
Figure \ref{figure_angle}.

It is obvious that the discretization (\ref{weight_method}) of $\Delta f$ can not
be a correct approximation of  $\Delta f$ since  it
approaches zero as the size of the surface mesh goes to zero.

\subsection{Mayer's et al. Discretization}
For a $C^2$ function $f$ on  $\Sigma$, Green's formula gives
 \begin{equation}\label{Mayer_weight}
\int_{D(z,\epsilon)} \Delta f(x) dx = \int_{\partial D(z,\epsilon)}
\partial_n f(s) ds
 \end{equation}

where $D(z,\epsilon)$ is a small disk at a point $z$ on the surface
$\Sigma$, and $n$ is the intrinsic outer normal of the boundary of
the disk. Mayer discretized (\ref{Mayer_weight}) at $v$ over the
triangular surface mesh $S$ and obtained the following
approximation

\begin{equation}\label{Mayer_method}
\Delta f(v) = \frac{1}{A(v)}\sum_{i \in N(v)} \frac{\|v_k-v_i\|
+ \| v_m - v_i\|}{2\|v-v_i\|}(f(v_i) - f(v))
\end{equation}
where $A(v)$ is the sum of areas of triangles around $v$, and
$k,m \in N(v) \cap N(v_i) $. It can be checked directly that the
formula (\ref{Mayer_method})  is derived from (\ref{Mayer_weight})
by approximating $\int_{D(z,\epsilon)} \Delta_{\Sigma} f(x) dx$, $\partial_n
f(s)$ and $ds$ with $\Delta f(v)A(v)$, $\frac{f(v_i)-
f(v)}{\|v_i - v\|}$ and $\frac{\| v_k - v_i\| + \|v_m -
v_i\|}{2}$, respectively. Therefore, the discretization in
(\ref{Mayer_method}) is an approximation of  $\Delta_{\Sigma} f$ at $v$.

\subsection{Desbrun's et al. Discretization}
It is well-known in the theory of differential geometry that the
mean curvature normal satisfies the following formula:
\begin{equation}\label{mean_curvature}
\lim_{diam(A) \rightarrow 0}\frac{3\nabla A}{2A} = -H(p)
\end{equation}
where $A$ is the area of a small region around the point $p$, and
$\nabla$  is the gradient with respect to the $(x,y,z)$ coordinates
of  $p$. From Equation (\ref{mean_curvature}), Desbrun et al. got the following
approximation:
\begin{equation}\label{Desbrun_method}
\Delta f(v) = \frac{3}{A(v)} \sum_{i \in N(v)} \frac{\cot
\alpha_i + \cot \beta_i}{2} |f(v_i) - f(v)|.
\end{equation}
where $N(v)$ is the index set of 1-ring neighboring vertices of
vertex $v$, $\alpha_i$ and $\beta_i$ are as in
(\ref{Desbrun_weight}) and $A(v)$ is the sum of areas of triangles
around $v$.

\subsection{Xu's Discretization}

In 2004, Xu presented two discrete Laplace-Beltrami methods in a triangular mesh from Green's formula and the quadratic fitting.
Following Equation (\ref{Mayer_weight}), xu introduced his discretization
\begin{equation}\label{Xu_method1}
\Delta f(v) = \frac{1}{2A(p)}\sum\limits_{i \in N(v)} <\nabla f(p) + \nabla f(p_i), \nu_i> \|v_i-v_{i_+}\|
\end{equation}
where $\nabla f(v)$ is the gradient of $f$ at $v$, $A(v)$ is the sum of area of the triangles that contain
$v$ and $\nu_i$ is the unit outward normal of the edge $\overline{v_iv_{i_+}}$. Xu also used the biquadratic fitting
 of the surface data and function data to calculate the approximate LB operator. He
introduced  complexity weights of the equation (\ref{weight_method}). This kind of weights can be found in \cite{Xu2}.

The convergence problem of these discrete LB operators over
triangular surface meshes has been investigated by Liu, Xu and Zhang in
\cite{Liu, Xu, Xu2}.  None of the above mentioned
discretizations of the LB operators has ever been proved to be
convergent for general surfaces and functions. The Desbrum et al.'s
discretization (\ref{Desbrun_method}) has been investigated under
some very restricted conditions. It is shown in \cite{Liu, Xu, Xu2}
that the discretization (\ref{Desbrun_method})  converges to the LB
operator under the conditions that the valence of the vertex $v$
is $6$ and $v = F(q)$, $v_i = F(q_i)$ for a smooth parametric
surface $F$  and the relations $q_{i+3} + q_i = 2q$, $i = 1,2,3$
hold, where $q_i$, $i=1,2,\cdots,6$ are one-ring neighboring
vertices of $q_i$ in the 2D domain. See \cite{Liu, Xu, Xu2} for more
details.

\section{A new convergent discrete algorithm for LB operators}
In this section we will describe a simple and effective method to
define the discrete LB operator on functions on a triangular mesh.
The primary ideas were developed in Chen, Chi and
Wu\cite{Chen3,Chen4} where we try to estimate  discrete partial
derivatives of functions on 2D scattered data points. Indeed, the
method that we  use to develop our algorithm is divided into
two main steps: First, we lift the 1-neighborhood points to the
tangent space and obtain a local tangential polygon. Second, we use
some geometric ideas to lift functions to the tangent space. We call
this a local tangential lifting (LTL) method. Then we present a new
algorithm to compute their Laplacians in the 2D tangent space. This
means that the LTL process allows us to reduce  2D curved
surface problems to  2D Euclidean problems. As one will see later,
our approach of discretization is quite different from the
discretizations discussed in section 2.

Consider a triangular surface mesh $S=(V,F)$, where $V=\{v_i| 1\leq
i \leq n_V \}$ is the list of vertices and $F=\{ f_k | 1 \leq k \leq
n_F\}$ is the list of triangles.

\subsection{ The local tangential lifting (LTL) method}
To describe the local tangential lifting (LTL) method, we introduce the approximating tangent plane $TS_A(v)$ and
the local tangential polygon $P_A(v)$ at the vertex $v$  of  as
follows:
\begin{enumerate}
\item The normal vector  $N_A(v)$ at the vertex $v$  in $S$  is given by
\begin{equation}
N_A(v) = \frac{\sum_{T \in T(v)}\omega_TN_T}{\|\sum_{T \in
T(v)}\omega_TN_T\|}
\end{equation}
where  $T(v)$ is the set of triangles that contain the vertex $v$,
$N_T$ is the unit normal to a triangle face $T$  and the centroid
weight is given in \cite{Chen,Chen2} by
\begin{equation}
\omega_T = \frac{\frac{1}{\|G_T - v\|^2}}{\sum_{\tilde{T} \in T(v)}
\frac{1}{\|G_{\tilde{T}} - v \|^2}},
\end{equation}
where $G_T$ is the centroid of the triangle face $T$determined by
\begin{equation}
G_T = \frac{v+v_i+v_j}{3}.
\end{equation}
Note that the letter $A$ in the notation $N_A(v)$  stands for the
word "Approximation".

\item The approximating tangent plane $TS_A(v)$  of $S$  at $v$   is now determined by
$TS_A(v) = \{ w \in \mathbb{R}^3 | w \bot N_A(v) \}$.

\item The local tangential polygon $P_A(v)$  of $v$ in $TS_A(v)$ is formed
by the vertices $\bar{v}_i$ which is the lifting vertex of $v_i$
adjacent to $v$  in $V$:
\begin{equation}
\bar{v}_i = (v_i - v) - < v_i - v,N_A(v)>N_A(v)
\end{equation}
as in figure \ref{figure_tangential_polygon}.

\begin{figure}[htb]
{\center
\includegraphics[width=6cm]{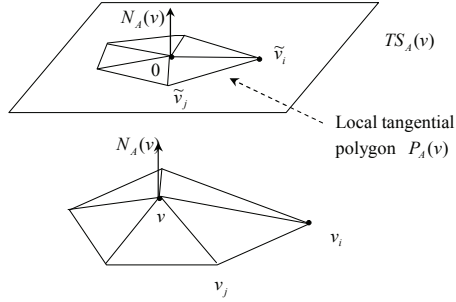}
\caption{The local tangential polygon $P_A(v)$}\label{figure_tangential_polygon}
}
\end{figure}

\item We can choose an orthonormal basis $e_1,e_2$ for the tangent plane
$TS_A(v)$ of $S$ at $v$ and obtain an orthonormal coordinates
$(x,y)$ for vectors $w \in TS_A(v)$ by $w = xe_1+ye_2$. We set
$\bar{v}_i = x_ie_1 + y_ie_2$ with respect to the orthonormal basis
$e_1, e_2$.

\end{enumerate}

Next we explain how to lift locally a function defined on $V$  to
the local tangential polygon $P_A(v)$. Consider a function  $h$ on
$V$. We will lift locally the function $h$ to a function of two
variables , denoted by $\bar{h}$, on the vertices $\bar{v}_i$ in
$P_A(v)$ by simply setting
\begin{equation}
\bar{h}(x_i,y_i) = h(v_i)
\end{equation}
and $\bar{h}(\vec{0}) = h(v)$  where  $\vec{0}$ is the origin of
$TS_A(v)$ . Then one can extend the function $\bar{h}$ to be a
piecewise linear function on the whole polygon  $P_A(v)$ in a
natural and obvious way.

\subsection{A new discrete 2D Laplacian algorithm and configuration equation}
In this section we present a new discrete 2D algorithm for
Laplacians acting on functions on the 2D domains in the $x-y$ plane.
Given a $C^2$  function $f$  on a domain $\Omega$  in the $x-y$
plane with the origin $(0,0) \in \Omega$, Taylor's expansion for two
variables x and y gives
\begin{equation}\label{Taylor_expansion}
f(x,y) = f(0,0) + xf_x(0,0) + yf_y(0,0) + \frac{x^2}{2}f_{xx}(0,0)+
xyf_{xy}(0,0)+\frac{y^2}{2} f_{yy}(0,0) + O(r^3)
\end{equation}
when $r= x^2+y^2$ is small.

Consider a family of neighboring points $(x_i,y_i) \in \Omega$.
$i=1,2, \cdots, n$, of the origin $(0,0)$. Take some constants
$\alpha_i$, $i=1,2, \cdots, n$, with $\sum\limits_{i=1}^n \alpha_i^2
= 1$. Then one has
\begin{equation}
\begin{array}{ll}
& \sum\limits_{i=1}^n \alpha_i( f(x_i,y_i) - f(0,0)) \cr =
&(\sum\limits_{i=1}^n \alpha_ix_i)f_x(0,0)+ (\sum\limits_{i=1}^n
\alpha_iy_i)f_y(0,0) + \frac{1}{2}(\sum\limits_{i=1}^n
\alpha_ix_i^2)f_{xx}(0,0) \cr + & (\sum\limits_{i=1}^n \alpha_i x_i
y_i) f_{xy}(0,0) + \frac{1}{2}(\sum\limits_{i=1}^n \alpha_i y_i^2)
f_{yy}(0,0) + O(r^3)
\end{array}
\end{equation}
We choose the constants $\alpha_i$, $i=1,2, \cdots, n$,  so that they
satisfy the following equations:

\begin{align}
\sum\limits_{i=1}^n \alpha_ix_i = 0, \tag{I}
\end{align}
\begin{align}
\sum\limits_{i=1}^n \alpha_iy_i = 0, \tag{II}
\end{align}
\begin{align}
\sum\limits_{i=1}^n \alpha_ix_iy_i = 0, \tag{III}
\end{align}
 and
$$\sum\limits_{i=1}^n \alpha_ix_i^2 = \sum\limits_{i=1}^n
\alpha_iy_i^2 $$ or equivalently
\begin{align}
\sum\limits_{i=1}^n \alpha_i(x_i^2- y_i^2) = 0, \tag{IV}
\end{align}

One can rewrite these equations in a matrix form and obtain the following the configuration equation:
\begin{equation}\label{alpha_matrix}
\begin{pmatrix}
x_1, & x_2, & \cdots, & x_n \cr y_1, & y_2, & \cdots, & y_n \cr
x_1y_1, & x_2y_2, & \cdots, & x_ny_n \cr

x_1^2 - y_1^2, & x_2^2 - y_2^2, & \cdots, & x_n^2-y_n^2 \cr
\end{pmatrix}\begin{pmatrix} \alpha_1 \cr \alpha_2 \cr \vdots \cr
\alpha_n \end{pmatrix} = \begin{pmatrix} 0 \cr 0 \cr 0 \cr
0\end{pmatrix}
\end{equation}
The solutions $\alpha_i$ of this equation allow us to obtain a formula for the Laplacian
$\Delta f(0,0)$:
\begin{equation}\label{laplace_1}
\begin{array}{ll}
\Delta f(0,0) & = f_{xx}(0,0) + f_{yy}(0,0) \cr\cr  & =
\frac{2\sum\limits_{i=1}^n
\alpha_i(f(x_i,y_i)-f(0,0))}{\sum\limits_{i=1}^n \alpha_i x_i^2}
+O(r)
\end{array}
\end{equation}
For the reason of symmetry, Equation (\ref{laplace_1}) also gives
\begin{equation}\label{laplace_2}
\Delta f(0,0)  = \frac{4\sum\limits_{i=1}^n
\alpha_i(f(x_i,y_i)-f(0,0))}{\sum\limits_{i=1}^n \alpha_i(
x_i^2+y_i^2) } +O(r)
\end{equation}
since we have $\sum\limits_{i=1}^n \alpha_i x_i^2 =
\sum\limits_{i=1}^n \alpha_i y_i^2$ . It is worth to point out that
Equation (\ref{laplace_2}) is an generalization of the well-known
5-point Laplacian formula. In the 5-point Laplacian case, we have
the origin $(0,0)$ along with 4 neighboring points $(s,0)$, ($0,s)$
$(-s,0)$ and $(0,-s)$ for sufficiently small positive number $s$.
From Equation (\ref{alpha_matrix}), one can find a solution
$\alpha_i = \frac{1}{2}$ for $i=1,2,3,4$, in this case.

\subsection{A new discrete approximation for LB operators over
surfaces}

 Now we can come back to handle the local lifting function $\bar{h}$
and propose a new discrete approximation for the LB operator over a
triangular surface mesh $S$  . From Equation (\ref{laplace_2}), we
can define a discrete LB operator $\Delta_A$ for the function $h$ at
the vertex $v$ by
\begin{equation}\label{laplace_h}
\Delta_A h (v) = \frac{4 \sum\limits_{i=1}^n \alpha_i(h(v_i) -
h(v))}{\sum\limits_{i=1}^n \alpha_i(x_i^2+y_i^2)}
\end{equation}
where the constants $\alpha_i$, $i=1,2,\cdots, n$ satisfy the configuration equation
(\ref{alpha_matrix}). Note again we have $\bar{v}_i = (v_i - v) -
<v_i - v, N_A(v)>N_A(v)$  and $\bar{v}_i = x_ie_1+y_ie_2$. Indeed,
this definition of the discrete LB operator  $\Delta_A$ is
independent of the choice of the orthonormal basis $e_1, e_2$. It
depends on the choice of the constants $\alpha_i$. To obtain a
unique solution $\alpha_i$ for $i=1,2,\cdots, n$, with $\sum_{i=1}^n
\alpha_i^2 = 1 $, one can simply choose $5$ closest neighboring points
 $\tilde{v}_i$ of the origin in $P_A(v)$. In case that the local polygon has less than $5$ vertices, one
can also lift the 2 or 3-ring of neighboring vertices of $v$ in $V$. In this way,
we can call $\Delta_A h(v)$ a 6-point Laplacian formula.

\subsection{ A convergence theorem for the discrete LB operators }
In this section we will prove that the discrete LB operator
$\Delta_A h$ for a smooth function  $h$ on a regular surface
$\Sigma$ is a convergent $O(r)$ approximation of the true LB
operator $\Delta_{\Sigma} h$. To show this, let $\Sigma$ be a smooth
regular surface in the 3D Euclidean space $\mathbb{R}^3$ and  $p \in
\Sigma$. Consider the exponential map $\exp_p : T\Sigma(p)
\rightarrow \Sigma$ from the tangent plane $T\Sigma(p)$ of $\Sigma$
at the point $p$ into the surface $\Sigma$. See do Carmo
\cite{Docarmo, Docarmo2} for discussions about the properties of the
exponential map $\exp_p$. One of the well-known properties of the
exponential map $\exp_p$  is that it is a local diffeomorphism
around the origin $\vec{0}\in T\Sigma(p)$. In other words, if $W$ is
a sufficiently small open domain around the origin  $\vec{0}$,
$\exp_p:W \rightarrow D$ is a diffeomorphism  where $D = \exp(W)$ is
an open domain around $p$. In particular, the inverse of $\exp_p$
exists on $D$.

Given a smooth function $h$ on the regular surface $\Sigma$, we can
lift $h$ via the exponential map $\exp_p$ locally to obtain a smooth
function $\hat{h}$ defined on $W \in T\Sigma(p)$  by setting
\begin{equation}
\hat{h}(w) = h(\exp_p(w))
\end{equation}
for $w \in W$. Fix an orthonormal basis $\tilde{e}_1, \tilde{e}_2$
 for the tangent space  $T\Sigma(p)$. This gives us a coordinate system on  $T\Sigma(p)$.
Namely, for $w\in W$ we have $w=x\tilde{e}_1+y\tilde{e}_2$ for two
constants $x$ and $y$. Without ambiguity, we can identify the vector
$w \in W$ with the vector $(x,y)$ with respect to the orthonormal
basis $\tilde{e}_1, \tilde{e}_2$. In this way, the function
$\hat{h}$ can also give us a smooth function  $\tilde{h}$ of two
variables $x$ and $y$ by defining
\begin{equation}\label{tilde_h}
\tilde{h}(x,y) = \hat{h}(w)
\end{equation}
for $w = x\tilde{e}_1+y\tilde{e}_2$. Using these notations, we will
prove

\begin{Lem}\label{Lemma1}
One has
 \begin{equation} \Delta_{\Sigma} h(p) = \Delta \hat{h}(0) =
\Delta \tilde{h}(0,0)
\end{equation}
\end{Lem}

\textbf{Proof:}

It is well-known that the LB operator $\Delta_{\Sigma} h(p)$  acting
on a smooth  function $h$ at a point $p$ can be computed from the
second derivatives of $h$  along any two perpendicular geodesics
with unit speed. See do  Carmo \cite{Docarmo2} for details. Indeed,
we consider the following two perpendicular geodesics with unit speed in
$\Sigma$ by using the orthonormal vectors $\tilde{e}_1,
\tilde{e}_2$:
\begin{equation}
c_i(t) = \exp_p (t\tilde{e}_i), ~~ i=1,2
\end{equation}
with $c_i(0) = p$  and $\frac{dc_i}{dt}(0) = \tilde{e}_i$. One has
\begin{equation}
\begin{split}
\Delta_{\Sigma} h(p) = & \frac{d^2}{dt^2}h(c_1(t))|_{t=0} +
\frac{d^2}{dt^2}h(c_2(t))|_{t=0} \\
= & \frac{d^2}{dt^2}\hat{h}(t\tilde{e}_1)|_{t=0} +
\frac{d^2}{dt^2}\hat{h}(t\tilde{e}_2)|_{t=0} \\
= & \Delta \hat{h}(\vec{0}) \\
= & \frac{\partial^2 \tilde{h}}{\partial x^2}(0,0) +
\frac{\partial^2 \tilde{h}}{\partial y^2}(0,0) \\
= & \Delta \tilde{h}(0,0)
\end{split}
\end{equation}
 \qed
\\

Next we consider a triangular surface mesh $S=\{V,F\}$ for the
regular surface $\Sigma$, where $V=\{v_i|1 \leq i \leq n_V\}$  is
the list of vertices and $F = \{T_k \ 1 \leq k \leq n_F \}$ is the list
of triangles and the mesh size is less than $r$. Fix a vertex $v$ in
$V$. For each face $T \in F$ containing $v$, we have
\begin{equation}
N_{\Sigma}(v) = N_T + O(r^2)
\end{equation}
where $N_{\Sigma} (v)$ is the unit normal vector of the true tangent
plane $T\Sigma (v)$ of  $\Sigma$ at $v$ and $N_T$ is the unit normal
vector of the face  $T$. Since the approximating normal vector
$N_A(v)$, defined in section 3.1 is a weighted sum of these
neighboring face normals  $N_T$, we have
\begin{Lem}\label{Lemma2}
One has
\begin{equation}\label{equ_lemma2}
N_{\Sigma} (v) = N_A{v} + O(r^2)
\end{equation}
\end{Lem}
Due to this lemma, the orthonormal basis $\tilde{e}_1, \tilde{e}_2$
for the tangent plane $T\Sigma(v)$ will give us an orthonormal basis
$e_1,e_2$ for the approximating tangent space  $T\Sigma_A(v)=\{ w
\in \mathbb{R}^3 | w \bot N_A(v) \}$ by the Gram-Schmidt process in
linear algebra:
$$ e_1 = \frac{\tilde{e}_1 -<\tilde{e}_1, N_A(v)>N_A(v)}{\|\tilde{e}_1 -<\tilde{e}_1,
N_A(v)>N_A(v)\|},$$

and
$$e_2= \frac{\tilde{e}_2 -<\tilde{e}_2, N_A(v)>N_A(v) - <\tilde{e_2},e_1>e_1}{\|\tilde{e}_2 -<\tilde{e}_2, N_A(v)>N_A(v) -
<\tilde{e_2},e_1>e_1\|}.$$ Logically speaking, one can first choose
an orthonormal basis  $e_1,e_2$ for the approximating tangent space
$TS_A(v)$ and then apply the Gram-Schmidt process to obtain an
orthonormal basis $\tilde{e}_1, \tilde{e}_2$ for the tangent plane
$T\Sigma(v)$. In either way, we always have by Lemma \ref{Lemma2}
the following relations.

\begin{Lem}\label{Lemma3}
One has
\begin{equation}\label{equ_lemma3}
\tilde{e}_i = e_i + O(r^2), ~~ i=1,2
\end{equation}
\end{Lem}
Consider a neighboring vertex $v_i$  of $v$  in $V$. For $r$  small
enough, we can use the inverse of the exponential map $\exp_p$   to
lift the vertex $v_i$ up to the tangent plane $T\Sigma(v)$ and
obtain

$$\tilde{v}_i = \exp^{-1}_v (v_i) \in T\Sigma(v)$$
 and
$$ \tilde{v}_i = \tilde{x}_i \tilde{e}_1 + \tilde{y}_i \tilde{e}_2$$
for some constants. As discussed in section 3.1, we can also lift
the vertex $v_i$ up to the approximating tangent plane $T\Sigma_A(v)$  and get

$$ \bar{v}_i = (v_i - v) - <v_i - v, N_A(v)>N_A(v)$$
and
$$ \bar{v}_i = x_ie_1 + y_ie_2$$
for some constants $x_i, y_i$. Then Lemmas \ref{Lemma2} and \ref{Lemma3} yield
\begin{Lem}\label{Lemma4}
One has
\begin{equation}
\left \{ \begin{split}
\tilde{x}_i & = x_i + O(r^2) \\
\tilde{y}_i & = y_i + O(r^2)
\end{split}\right.
\end{equation}
\end{Lem}
Using these relations, one can solve the configuration equation (\ref{alpha_matrix}) for $(\tilde{x}_i,\tilde{y}_i)$
and $(x_i,y_i)$ respectively and obtain their corresponding solutions $\tilde{\alpha}_i$ and $\alpha_i$ with the relation
\begin{equation}
\tilde{\alpha}_i = \alpha_i + O(r^2)
\end{equation}
Note that the lifting function $\tilde{h}$ is a smooth function of two variables $x$ and $y$.
Equation (\ref{laplace_2}) in section 3.2 now gives an approximation of the Laplacian $\Delta \tilde{h}(0,0)$:
\begin{equation}
\Delta \tilde{h}(0,0) = \frac{4\sum\limits_{i=1}^n \tilde{\alpha}_i(\tilde{h}(x_i,y_i) -
\tilde{h}(0,0))}{\sum\limits_{i=1}^n\tilde{\alpha}_i( \tilde{x_i}^2+\tilde{y_i}^2)} + O(r)
\end{equation}
The relations (\ref{tilde_h}), (\ref{equ_lemma2}) and (\ref{equ_lemma3}) imply
\begin{equation}
\Delta \tilde{h}(0,0) = \frac{4\sum\limits_{i=1}^n \alpha_i(h(v_i) -
h(v))}{\sum\limits_{i=1}^n\alpha_i( x_i^2+y_i^2)} + O(r).
\end{equation}
This along with Lemma \ref{Lemma1} proves the following  convergence theorem.

\begin{MainThm}
Given a smooth function $h$ on a regular surface $\Sigma$  and a triangular surface mesh $S=(V,F)$  with mesh size $r$, one has
\begin{equation}
\Delta_{\Sigma} h(v) = \Delta_A h(v) +O(r)
\end{equation}
where the discrete LB operator $\Delta_A h(v)$ is defined by Equation (\ref{laplace_h}).
\end{MainThm}
\begin{Rem}
The discussions in this section also indicate that as long as we have $O(r)$-convergent algorithms to estimate gradients,
Laplacians and other intrinsic derivatives of 2D smooth functions, the LTL method and methods in section 3.4 will allow us
 to develop corresponding discrete convergent algorithms over 3D surfaces. It is possible to obtain a $O(r^2)$ algorithm by extending
  Taylor's expansion (\ref{Taylor_expansion})to the third order and improving the
 configuration equation (\ref{alpha_matrix}).
\end{Rem}

\section{Numerical simulations}

In this section, we shall compare two convergent Laplace-Beltrami methods: Xu's method (see \cite{Xu2})
and our proposed method. We take four functions,
$$ \begin{array}{ll}
F_1(x,y) &= \sqrt{4-(x-0.5)^2-(y-0.5)^2}. \cr\cr
F_2(x,y) &= \tanh(9x-9y). \cr\cr
F_3(x,y) &= \frac{1.25+\cos (5.4y) }{6+6(3x-1)^2}. \cr\cr
F_4(x,y) &= \exp \left ( -\frac{81}{16} ((x-0.5)^2+(y-0.5)^2) \right ).
\end{array}
$$
over $xy$-plane as three dimensional surfaces.

\begin{figure}[htb]
\includegraphics[width=13cm]{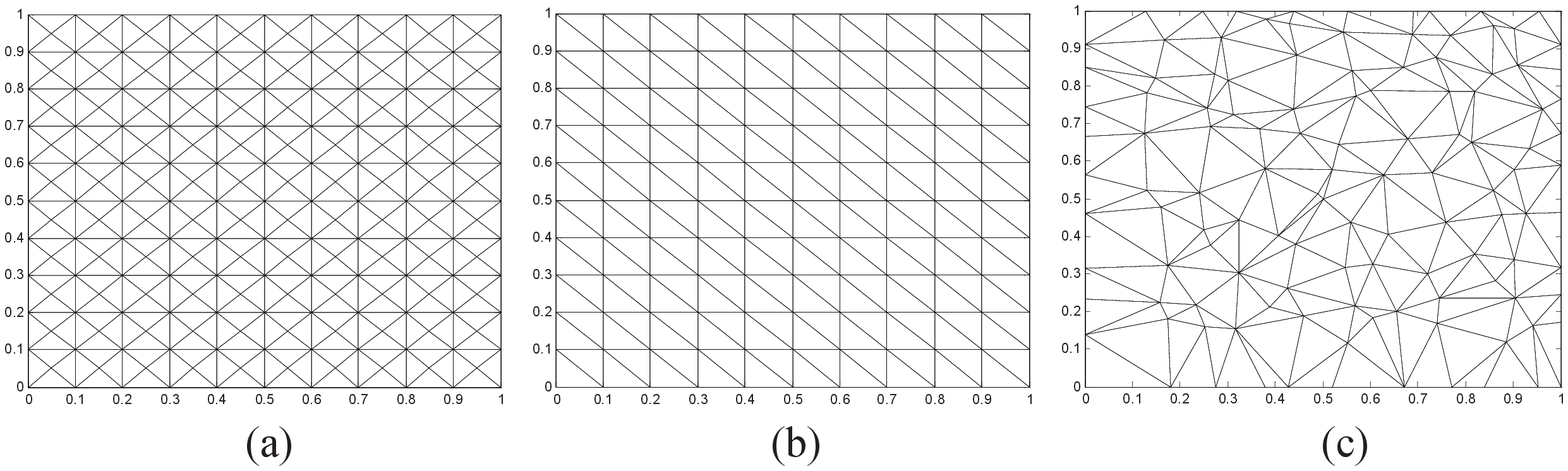}
\caption{The triangulation of the domain.}

\end{figure}

 The exact and approximated mean curvatures are computed as some selected interior
domain points $(x_i,y_j)$ with $x_i, y_j \not\in \{0,1\}$. The domain (a) is a three directional triangular partition,
the domain (b) is a four directional triangular partition and the domain (c) is a unstructured triangular partition. To observe the convergence
property, these domains are recursively subdivided by the bisection linear subdivisions. Hence, $h=\frac{h_0}{2^i}$, $i=1,2,\cdots$, where
$h_0 = \sqrt{0.2}, \frac{0.1}{\sqrt{2}}, 0.23$ are the maximal value of edge lengths of the triangulations (a), (b) and (c), respectively.

The maximal errors of these simulations are shown in the Table \ref{result_table1}. Table \ref{result_table2} shows
the time costs for the computations in the domain (c) with $h=\frac{h_0}{2^{10}}$. Obviously, our proposed method is more accurate and faster than
Xu's  method. Furthermore, the convergent rate of our  method is also better than   Xu's method.

\begin{table}[htb]
\caption{The maximal errors of Laplacian}\label{result_table1}
{\centering
\begin{tabular}{lllll}
\hline
&  & & Xu's method & Our method \\
\hline
            &\vline& $F_1$ & $2.34E-03*h^2$  &  $2.36E-05*h^4$ \\
Domain (a)  &\vline& $F_2$ & $1.81E+01*h^2$  &  $3.15E-02*h^3$ \\
            &\vline& $F_3$ & $9.15E-01*h^2$  &  $1.58E-02*h^3$ \\
            &\vline& $F_4$ & $1.50E-+1*h^2$  &  $1.68E+00*h^3$ \\

\hline
            &\vline& $F_1$ & $4.24E-03*h^2$  &  $2.64E-05*h^2$ \\
Domain (b)  &\vline& $F_2$ & $1.89E+01*h^2$  &  $5.13E-01*h^2$ \\
            &\vline& $F_3$ & $1.47E+01*h^2$  &  $4.29E-01*h^2$ \\
            &\vline& $F_4$ & $3.51E+01*h^2$  &  $7.84E-01*h^2$ \\

\hline
            &\vline& $F_1$ & $1.49E-01$  &  $1.53E-03*h^2$ \\
Domain (c)  &\vline& $F_2$ & $1.94E+00*h^{\frac{1}{2}}$ &  $1.59E-01*h^2$ \\
            &\vline& $F_3$ & $9.22E-01$  &  $3.00E-01*h$\\
            &\vline& $F_4$ & $5.41E-01*h$ & $1.36E+01*h^2$ \\
\hline
\end{tabular}
}
\end{table}

\begin{table}[htb]
\caption{Time costs for the computations of domain (c)  }\label{result_table2}

{\centering
\begin{tabular}{lll}
\hline
(seconds)  & Xu's method & Our method \\
\hline
$F_1$ & $0.024$  &  $0.010$ \\
$F_2$ & $0.026$  &  $0.014$ \\
$F_3$ & $0.026$  &  $0.015$ \\
$F_4$ & $0.025$  &  $0.012$ \\
\hline
\end{tabular}

}

\end{table}

\section*{Acknowledgements}

This paper is partially supported by NSC, Taiwan.

\end{document}